# Low Contrast Dielectric Metasurface Optics

*Alan Zhan[1], Shane Colburn[2], Rahul Trivedi[3], Taylor K. Fryett[2], Chris M. Dodson[2], and Arka Majumdar[1,2,+]*

[1] *Department of Physics, University of Washington, Seattle.*

[2] *Department of Electrical Engineering, University of Washington, Seattle.*

[3] *Department of Electrical Engineering, Indian Institute of Technology, Delhi.*

[+] *Corresponding Author: arka@uw.edu*

**Abstract**:

The miniaturization of current image sensors is largely limited by the volume of the optical elements. Using a sub-wavelength patterned quasi-periodic structure, also known as a metasurface, one can build planar optical elements based on the principle of diffraction. However, recent demonstrations of high quality metasurface optical elements are mostly based on high refractive index materials. Here, we present a design of low contrast metasurface-based optical elements. We validate our theory by fabricating and experimentally characterizing several silicon nitride based lenses and vortex beam generators. The fabricated lenses achieved beam spots of less than 1 μm with numerical apertures as high as ~ 0.75. A transmission efficiency of 90% and focusing efficiency of 40% in the visible regime was observed. Our results pave the way towards building low loss metasurface-based optical elements at visible frequencies using low contrast materials, and extend the range of prospective material systems for metasurface optics.

**Main Text:**

Conventional transmissive macroscopic optical elements primarily depend on refraction to control the propagation of light. Refraction relies heavily upon the exact curvature of the surface, and the spatial extent of the element in order to achieve gradual phase accumulation. This imposes a fundamental limitation on the miniaturization of optical sensors and elements, which is necessary for various applications such as the Internet of Things[1], bio-photonics[2,3] and two photon absorption microscopy[4]. Metasurfaces, two-dimensional quasi-periodic arrays of sub-wavelength structures, present a novel method of miniaturizing optical elements. Rather than relying on gradual phase accumulation through light propagation, each sub-wavelength structure imparts a discrete, abrupt change in the phase of incoming light[5-7]. This has motivated the design of metasurface-based optical elements including lenses[8,9], focusing mirrors[10], vortex beam generators[11,12], holographic masks[13,14], and polarization optics[15,16].

Thus far, high quality metasurface optical elements based on metals[5,17], titanium oxide[18,19], and amorphous silicon[20,21] have been demonstrated. Unfortunately, metals are significantly lossy at optical frequencies[22], titanium oxide lacks CMOS compatibility, and amorphous silicon absorbs light in the visible and near-infrared spectrum (~400-900nm). This wavelength range is of particular interest for many applications due to ubiquitous, low-cost silicon detectors, motivating the development of high band gap material based metasurfaces. However, high band gap CMOS-compatible materials such as silicon nitride and silicon dioxide, which are transparent over the aforementioned wavelength range, have a low refractive index. Although silicon dioxide metasurface lenses have been previously demonstrated, they had low numerical apertures, resulting in large beam spots[23,24]. In this paper, we demonstrate operation of high quality metasurface lenses (NA~0.75) and vortex beam generators based on silicon nitride at visible wavelengths. Our results present a methodology for producing low-loss high quality metasurface optics that is compatible with both silicon detectors, and conventional CMOS fabrication technologies.

The main building block of a metasurface is a grating composed of scatterers arranged in a sub-wavelength periodic lattice (with a period p). In this paper, we focus on cylindrical posts as the scatterers, which are arranged on a square lattice (Figure 1a). For such a grating, the higher order diffracted plane waves are evanescent and only the zeroth order plane wave propagates a significant distance from the grating[25]. The complex transmission coefficient of this plane wave depends upon the grating periodicity $p$, scatterer dimensions (both the diameter $d$ and thickness $t$), and refractive index $n$, as shown in Figure 1. The use of metasurfaces to build optical components is primarily motivated by the observation that the functionality of many such components, such as lenses and focusing mirrors is determined by a spatial phase profile imparted on an incident beam. Reproducing these devices using a metasurface involves selecting the correct parameters to achieve the desired spatial phase profile, arranging the scatterers on a sub-wavelength lattice, and spatially varying their dimensions. In order to design an arbitrary transmission phase profile, however, we must be capable of producing phase shifts spanning the whole 0 to $2\pi$ range while maintaining large transmission amplitudes. Such phase variations have been demonstrated earlier with high refractive index scatterers (Figure 1b). Via numerical simulation using rigorous coupled wave analysis (RCWA)[26], we found that it is possible to select grating parameters so as to achieve such a phase variation with a low contrast grating. In these simulations, we calculate the achievable phases and transmission amplitudes by varying the diameter $d$ of the posts for a fixed periodicity $p$, substrate thickness ($t_{sub} = \lambda$), and refractive index of n~2. At first glance, one might assume that one can arrive at a low contrast grating refractive index of $n$~2. A possible procedure for arriving at a low contrast grating design is by simply scaling the high contrast grating inversely with the grating index. However, while such a simple scaling method can produce phase shifts spanning the whole 0 to $2\pi$ range, it also results in severe dips in the transmission amplitude due to the appearance of resonances (Figure 1c). Note that, these resonances also appear in the high contrast case (Figure 1b), but they are generally significantly narrower when compared to the low contrast case. The broad resonances in the low contrast design result in strongly varying transmission amplitudes, rendering it unsuitable for an efficient optical element.

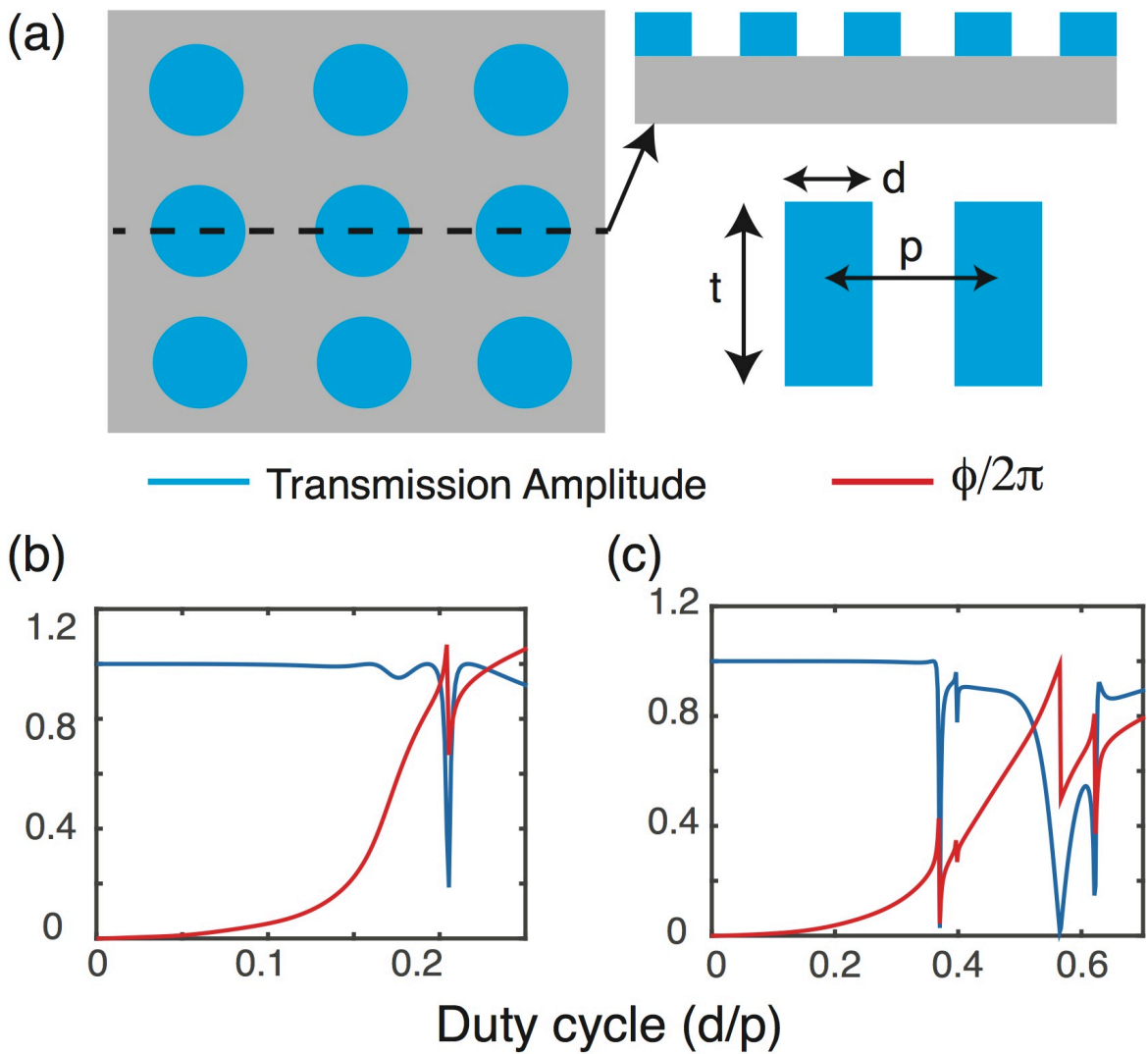

Figure 1: Low and high contrast metasurfaces: (a) Schematic of the grating structures: a grating with periodicity p can be formed by using cylindrical posts (with diameter d) arranged in a

*square lattice. The thickness of the grating is denoted as t. Amplitude and phase ϕ of the transmitted light for (b) a high contrast ($n_{high}$ = 3.5) and (c) a low contrast ($n_{low}$ = 2.0) grating using parameters from (b) scaled by $n_{high}/n_{low}$.*

These resonances can however, be engineered by choosing different grating parameters, such as thickness and periodicity. Specifically, by varying the thickness and periodicity of the low contrast grating, we can transition from a region with many resonances to a non-resonant regime. Simulation results with varying thicknesses and periodicities are shown in Figure 2. For $t = 1.2\lambda$ and $p = 0.4\lambda$, the phase delay and transmission amplitude are both continuous for all the post-diameters, with only small variations in the transmission amplitude (Figure 2g). This set of parameters can be considered to be fully within the non-resonant regime as it lacks any discontinuities in the phase or transmission amplitude. Unfortunately, this design has a large aspect ratio, making it difficult to fabricate. By increasing the thickness of the pillars, the resonances are narrowed for a given periodicity (Figure 2c,i). Additionally, increasing the periodicity for a fixed thickness results in more resonances in all cases. Based on these simulations, we chose the parameters $t = \lambda$ and $p = 0.7\lambda$ (Figure 2h), to ensure a moderate aspect ratio for fabrication while maintaining near unity transmission amplitudes for the whole range of phases.

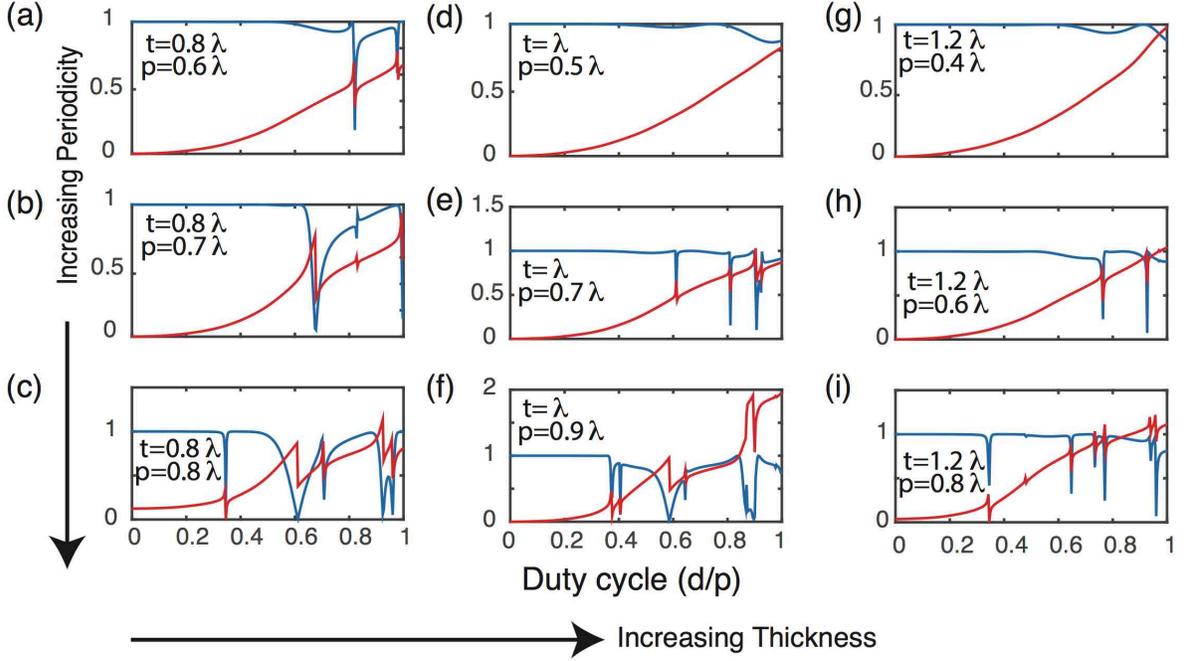

*Figure 2: Phase and transmission behavior for low contrast metasurfaces with different diameters, periodicities, and thicknesses: Phase delay (red) and transmission amplitude (blue) for gratings as a function of the duty cycles for varying periodicity and thickness.*

We can realize any arbitrary phase profile by arranging these scatterers in a lattice. We chose to fabricate aspheric lenses and a vortex beam generator due to their relatively simple phase profiles. The spatial phase profile of a lens is given by:

$$\Phi(x, y) = \frac{2\pi}{\lambda}\left(\sqrt{x^2 + y^2 + f^2} - f\right), \qquad (1)$$

where $f$ is the focal length of the lens, $(x,y)$ are the in-plane coordinates, $z$ is the propagation direction, and $\lambda$ is the design wavelength. The spatial phase profile of a focusing vortex beam generator is:

$$\Phi(x, y) = \frac{2\pi}{\lambda}\left(\sqrt{x^2 + y^2 + f^2} - f\right) + \ell\theta, \qquad (2)$$

which is a lens modified by the angular momentum term ($\ell\theta$), where $\ell$ is an integer specifying the orbital angular momentum state and $\theta$ is the azimuthal angle in the lens plane. In our design, we map the spatial phase profile onto a square lattice by discretizing the phase profile into six steps. For each discrete value of the phase profile, we find the radius of the pillar that most closely reproduces that phase. In our lens, the radii of the pillars vary from 96 nm to 221 nm, all with thickness equal to $\lambda$ (633 nm) on a lattice with a periodicity of $0.7\lambda$ (443 nm) corresponding to Figure 2e.

To validate our theory, we fabricated and characterized metasurface lenses and vortex beam generators in silicon nitride (n ~ 2). Figures 3a and 3b show respectively, a scanning electron micrograph (SEM) of the fabricated lens and vortex beam generator. We prepared the wafer by depositing 633 nm of silicon nitride on a 500 μm thick quartz substrate using plasma enhanced chemical vapor deposition (PECVD). 50 nm of aluminum was then evaporated onto the silicon nitride, serving both as a hard mask and as a charge-dissipation layer necessary for electron beam lithography. The pattern was exposed on 160 nm ZEP 520A using a Jeol JBX-6300FS 100 kV electron beam lithography system. Following development in amyl acetate, the sample was dry etched with a $Cl_2$ and $BCl_3$ plasma to transfer the pattern on the aluminum layer, forming the hard mask. Finally, a $CHF_3$ and $O_2$ plasma was used to etch the 633 nm pillars, and the remaining aluminum was removed using sulfuric acid.

The schematic of the optical setup used to probe the structures is shown in Figure 3c. The characterization setup utilizes a 40x objective (Nikon Plan Fluor) with a working distance of 0.66 mm and NA 0.75, and a tube lens (Thorlabs SM2A20) with a focal length of 20 cm as a microscope. The magnification of the setup was determined using known dimensions of the lens. We mount the metasurface on a glass slide with the front facing the microscope. The devices were illuminated with red (Thorlabs M625F1), green (Thorlabs M530F1), and blue (Thorlabs M455F1) light-emitting diodes (LEDs).

The intensity profiles were captured using the microscope and a Point Grey Chameleon camera. By translating the microscope and camera along the optical axis, we can move into and out of the focal plane and image the x-y plane intensity profile at varying z-distances. During characterization we can clearly see that the beam radius is changing as we translate into and out of the focus. The full-width half maxima (FWHM) values obtained by a Gaussian fit are plotted in Figure 4a as a function of the distance in the z-direction. The increasing error bars away from the focal distance are due to the divergence of the single peak into two peaks as the microscope moves out of the focal plane. A typical example of the intensity profile near the focal point is shown in Figure 4b. A Gaussian fit is shown in Figure 4c.

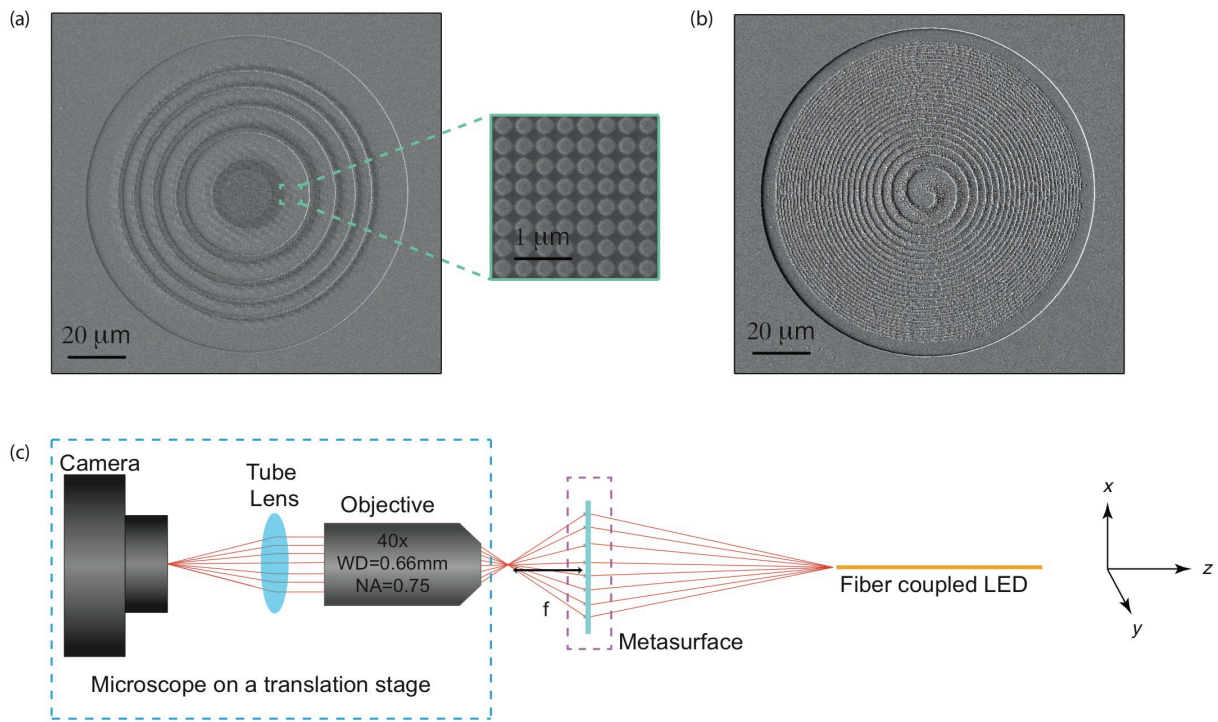

*Figure 3: Metasurface lens and setup: (a) Scanning electron micrograph (SEM) of a f = 0.5 mm lens. Zoom shows aluminum-capped silicon nitride pillars arranged on a square lattice. (b) SEM of a focusing vortex beam generator with ℓ = 1 and f = 100 µm. (c) Microscopy setup for imaging the focal plane. The microscope can be translated along the optical axis.*

We fabricated lenses with five different focal lengths between 50 µm and 1 mm with a lens radius of 56 µm. The measured FWHM of the focal spot sizes for all the lenses are plotted against the ratio of the lens focal length ($f$) to lens diameter ($d$) in figure 5a, where the dotted green line is the FWHM of a diffracted limited spot of a lens with the given geometric parameters. The deviation from the diffraction limit is attributed mostly due to fabrication imperfection. The criterion for the diffraction limited FWHM spot is given in supplement S1, and simulated spot sizes for low $f/d$ lenses are provided in supplement S2. The 50 µm lens achieves a spot size diameter of less than 1 µm. The measurement of the focal distance also agrees well with our design parameters.

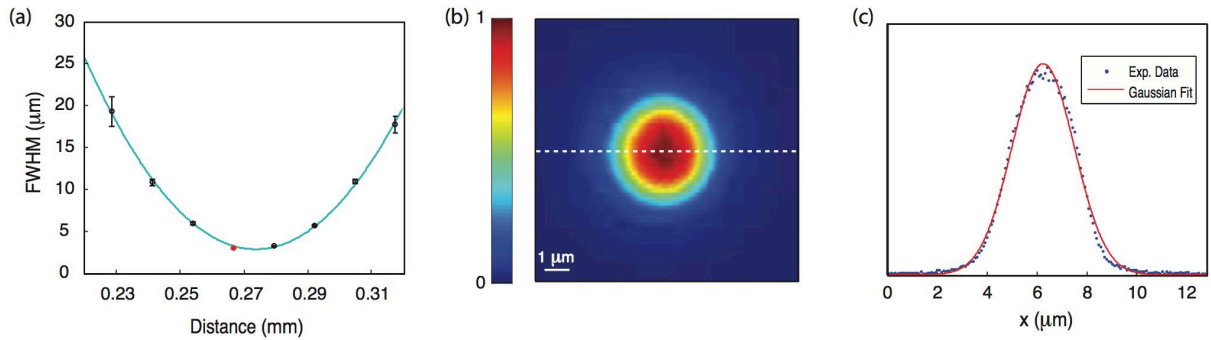

*Figure 4: 250 µm focal length lens performance measured with a LED centered at 626 nm: (a) The FWHM is plotted as a function of the distance in the z-direction. The working distance of the objective has been subtracted. The error bars denote the 95% confidence interval for the Gaussian fits. The blue curve is an eye guide. (b) 2D intensity profile at the focal plane, the red point in (a). (c) A Gaussian function is fit to the cross-section data to estimate the beam size.*

*Cross-section taken from dashed white line in (c). We use the FWHM as a measure for the beam size.*

In order to measure the focusing efficiency of the lens we inserted a flip mirror before the camera to direct the beam to a power meter (Newport 1918-R). We then measure the incident power to the focus by using a pinhole to isolate a spot with radius three times the FWHM. The focusing efficiency was taken to be the ratio of the power incident on the focus to the power incident on the lens. The transmission efficiency was taken to be the ratio of the power incident on the detector through the lens to the power incident through a glass slide. Transmission and focusing efficiencies both show an increase as the focal length of the lens increases as shown in Figure 5b. The focusing efficiency reaches a maximum of ~40% for the 1 mm lens and the transmission efficiency rises to near 90% for the 500 μm lens. These transmission efficiencies are significantly higher than other metasurfaces in the visible frequency range[14, 21, 27]. Simulated efficiencies for both low and high contrast lenses with low $f/d$ are provided in supplement S2. In addition, we investigated the chromatic behavior of the lens for red, green, and blue light. The wavelength dependence of the 250 μm focal length lens is shown in figure 5c. The focal distance of our lens increases with decreasing wavelength, increasing from ~0.26 mm at 625 nm to ~0.35 mm at 455 nm. We also observe an increase in the size of the focal spot with decreasing wavelength, from a minimum of ~3 μm at 625 nm to a maximum of ~4 μm at 455 nm. We remark that the product of the experimentally measured focal length ($f$) and illumination wavelength ($\lambda$) is roughly constant for our design, as expected[28].

Finally, we characterized the vortex beam generators and imaged their intensity profiles as shown in Figure 6a and b. They were fabricated with a focal length of 100 μm, radius of 60 μm, and orbital angular momentum states ℓ = 1 (a), and ℓ = 2 (b). The experimentally measured spiral intensity profiles clearly show the two distinct orbital angular momentum states.

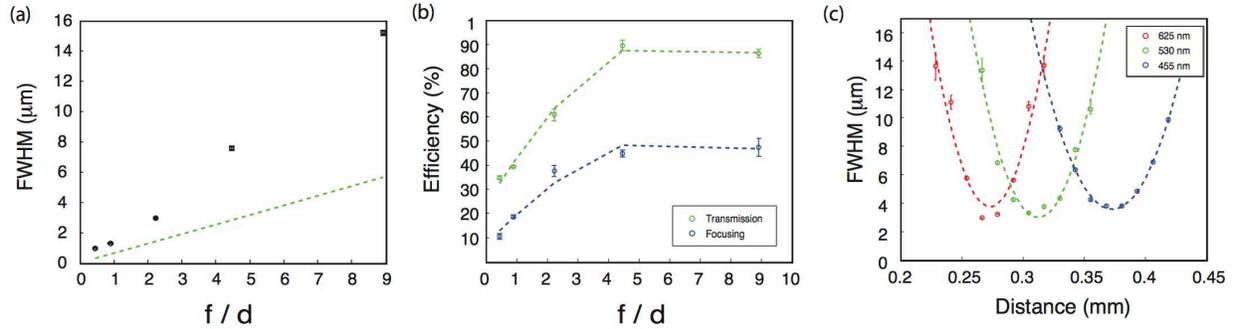

*Figure 5: Performance of metasurface lenses plotted as a function of the ratio of their focal length to diameter: (a) Measured focal spot sizes for all fabricated lenses. The dashed green line is the diffraction limited FWHM. The justification for the green line is shown in supplement S1. (b) Measured transmission and focusing efficiencies for all fabricated lenses. Error bars are obtained from the standard deviation of three measurements on each device. (c) Chromatic dispersion of the lens. Red, blue, green correspond to illumination with 625 nm, 530 nm, and 455 nm LEDs respectively. Plotted curves are eye guides, and error bars represent the 95% confidence intervals for the Gaussian fits.*

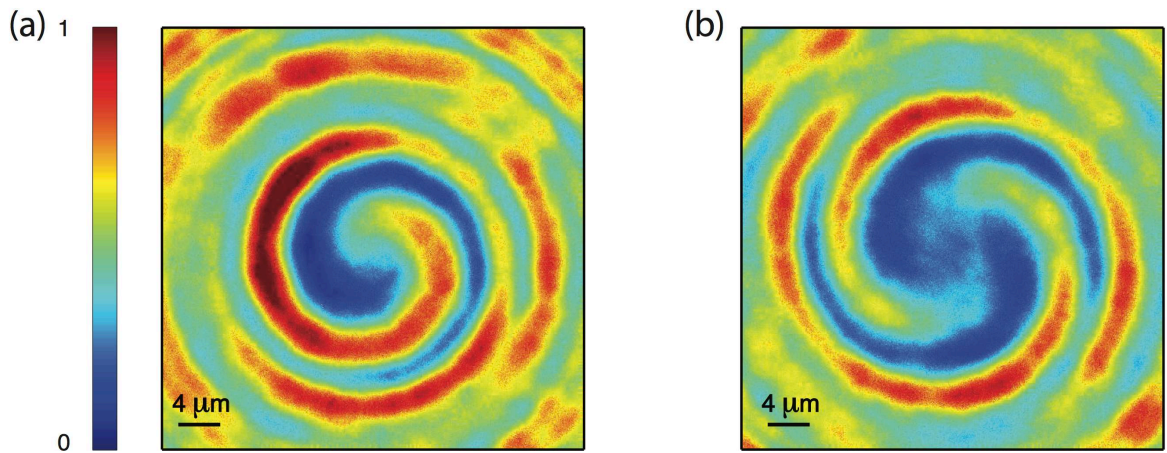

*Figure 6: Vortex beam generator normalized intensity field profiles for (a) ℓ = 1, and (b) ℓ = 2 showing the distinct helical wave fronts. Both figures share the same color bar.*

We have designed and fabricated low contrast high quality metasurface optical elements based on silicon nitride. Our lenses achieved transmission efficiencies of up to 90% and focusing efficiencies of up to 40%, in addition to a sub 1 µm spot size, with a numerical aperture of 0.75. The performance of these lenses is significantly better than previously reported results in the context of low contrast diffractive gratings. Recent demonstration of optical elements with similar performance employed high contrast materials such as metals and silicon that are incompatible with operation in the visible spectrum. The formalism we developed is applicable for the design of arbitrary spatial phase profiles in the same way as previous designs, and brings the field of metasurface optics fully into the visible spectrum. Additionally, by using silicon nitride as our metasurface material, we can leverage both CMOS compatibility and low visible absorption to design our lenses. We emphasize that our analysis is not limited to silicon nitride (n ~ 2), but is also applicable to other low index materials such as transparent conducting oxides (TCO), organic polymers, transparent printable materials, and silicon dioxide. These materials, specifically TCOs and organic polymers, may provide an easier way to tune the metasurface elements due to their stronger electro-optic properties[29-32] or strong free carrier dispersion[33].

In conclusion, we have demonstrated a low contrast metasurface design allowing for the arbitrary shaping of an optical wavefront in the visible regime. The use of low contrast materials extends the range of materials available for metasurface optics. The wavelength-scale thickness and planar geometry of the optical elements allows the miniaturization of optical elements for integration on optical fibers for bio-photonics and use in small-scale optical systems. In addition, this approach greatly simplifies the design and fabrication process of complicated aspherical optical elements, including free-form optics[34].

**Acknowledgements:**

We would like to thank Dr. Andrei Faraon for useful discussions. All of the fabrication was performed at the Washington Nanofabrication Facility (WNF), a National Nanotechnology Infrastructure Network (NNIN) site at the University of Washington, which is supported in part by the National Science Foundation (awards 0335765 and 1337840), the Washington Research Foundation, the M. J. Murdock Charitable Trust, GCE Market, Class One Technologies and Google. The research work is supported by the startup fund provided by University of Washington, Seattle, and Intel Early Career Faculty Award.

**Associated Content:**

Supporting Information:

Figures S1, S2, S3, a criterion for diffraction limited lenses, FDTD simulations of achievable spot sizes and efficiencies for low and high contrast lenses. This content is available free of charge through the internet at http://pubs.acs.org.

**Table of Contents Figure:**

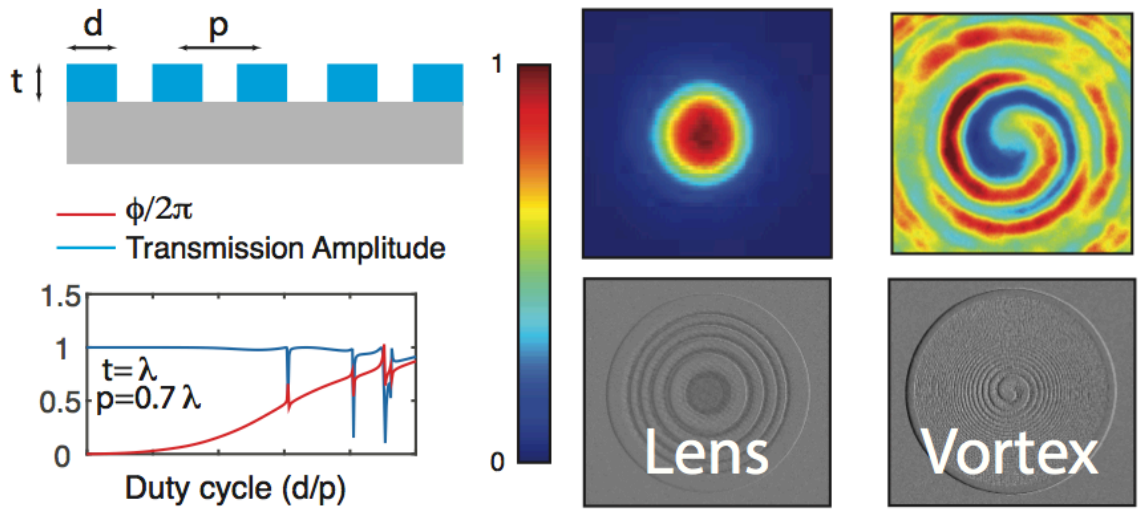